# Monitoring Open Access at a national level: French case study

Eric Jeangirard

## INTRODUCTION

1  The French National Plan for Open Science was launched in July 2018. The steering of Open Science is one of its pillars. Building an Open Access monitor for France is then an essential tool to fulfill this objective.

2  Before the start, we had several requirements for this tool: the methodology had to be transparent and the results reproducible. Plus, the outcome had to be granular so that multiple dimensions could be analyzed. Open Access (OA) trends can actually hide more subtle dynamics in different scientific fields for example.

3  As a consequence, that tool has to be based on open data or publicly available data on the Web to guarantee full transparency and reproducibility of the outcome. We present the methodology that we developed and tested for France, but that could be extended to other countries. We leverage on multiple existing initiatives, some international, some national, such as Unpaywall, HAL (main open repository in France), ORCID (Open Researcher and Contributor ID) and IDRef. IDRef is a referential for Higher Education and Research, built and maintained by the ABES (Bibliographic Agency of Academic Libraries which is French public institution and supervised by the Ministry for Higher Education Research and Innovation).

4  Our key objective is to measure and describe, on a yearly basis, the open access level for all the publications with at least a French affiliation. This objective raises 3 challenges: (1) identify the publications with a French affiliation, (2) enrich the metadata of the publications to make it as interoperable as possible and (3) detect the OA status of each publication.

5   In this work, we focus on a five-year period, from 2013 to 2017. For this period, we estimate the global OA rate. We also describe the OA variation between disciplines and publishers. We describe each step of the method and estimate the associated error rates.

6   Several attempts at building a monitor for Open Science and/or Open Access already exists.
The European Commission launched an Open Science Monitor initiative but chose to use "closed" data (Scopus). That makes this methodology not relevant for our use case as we want it to be fully transparent and reproducible.

7   A recent paper from Piwowar, Priem et al. (2018) also analyses the OA trends using oaDOI service (previous name of Unpaywall). But this is a global approach that cannot be used directly for national level as geographical metadata are most of the time missing.

8   At a national or regional level, multiple works are ongoing. For example, for Catalonia, Rovira and Labastida (2018) set up a methodology using the Wos and Scopus to get the publications and then Unpaywall to get the OA status.

9   Countries like the Netherlands, Denmark, and Germany also work on national Open Access monitoring. Some of them rely on a bottom-up data flow. In Denmark for example, publication metadata is collected from each of the local research databases of the 8 Danish universities.

10  None of the existing approaches allowed to track the OA rate at a national level without a bottom-up data production system or the use of an existing database that already has geographical metadata. Our methodology instead relies on huge open databases and referential and uses simple (for matching) and more complex algorithms to enrich and filter the data.

11  We use a three-step methodology to build our Open Access Monitor. First, identifying which publications should be considered into the perimeter study (publications with at least a French affiliation). Second, enriching the metadata of the publications in the perimeter (author identification and discipline classification). Third and eventually deciding on the OA status of the publications in the perimeter.

## METHOD

### First step in the method: Identify the publications with a French affiliation

12  In this work, we decided to focus only publications with a DOI (Digital Object Identifier). We use data from Unpaywall and HAL (open repository) to list all the publications published during the period year. There are about 4.5 million DOIs per year. These sources provide also a few metadata for the publications: title, authors first name and last name, journal name, and International Standard Serial Number (ISSN). However, the affiliations metadata are generally missing (especially from Unpaywall, which is the main source we use).

13  As a consequence, to decide whether a given DOI has a French affiliation, we first check if it is available on HAL with a French affiliation. If not, we scrap the DOI redirect page, i.e http://doi.org/ followed by the DOI we want to analyze. From the HTML we get, we designed a parser (actually quite a diversity of specific and more generic parsers) that try to locate the affiliations in the HTML and decide if one of them is French. This task is actually a process automation of what a human would do to know if there is a French affiliation, that is looking for the DOI on the Web and then reading the affiliations on the article page.

14  We had to face several difficulties during this step.

1. The first challenge is the wide variety of web pages encountered so that we had to implement around 30 parsers to detect the affiliations in the page. Regular expressions (regex) is then used to determine if there is a French affiliation or not. These regex contain keywords like "France" and the main French cities names.

2. The second challenge is the volume, around 4.5 million of DOI per year. Scraping web pages should not be too aggressive, so that can be quite time-consuming. We decided to pre-filter the DOI and avoid to check all of them. This pre-filtering is based on authors first name and last name. We built a list of authors names for persons that, at some point in their career, were affiliated in a French structure. We used multiple data sources, in particular, *theses.fr (search engine for PhD defended in France)*, *Unpaywall* (some DOI entries do have affiliation metadata), *HAL, ORCID, Pascal & Francis (bibliographic databases maintained by INIST - Institut de l'Information Scientifique et Technique, CNRS)*. In this list, the same person can occur several times with different names form (e.g maiden name) and conversely, a unique name can represent several persons (homonyms). This pre-filtering is actually optional and was only a way to speed up our processes.

15  The affiliations metadata are key for building an OA monitoring at a national level. The method we used can be error prone and its accuracy has to be estimated.

16  Two types of errors can occur:

- False positive (precision), i.e a DOI is detected as having a French affiliation, but it has not. This happens when the rule-based detection is mistaken. This can be measured manually, with a check of a random sample.
- False negative (recall), i.e a DOI is not detected as having a French affiliation, but it actually has. This is much more difficult to measure as it is the 'invisible' part.

## Second step in the method: Enrich the metadata of the identified publications

17  Once the perimeter of the publications is defined after the first step, it is essential to describe them, that is to say, being able to answer these questions: who are the authors of these publications? in which structures do they work? to which scientific field do they belong?

18  For authors and affiliations, that means improving data interoperability by adding external identifiers from national and international databases and referentials like ORCID, IDRef, HAL for authors, and grid (Global Research Identifier Database), RNSR (*r épertoire national des structures de recherche,* which is the referential for French research structure maintained by the French Ministry of Higher Education and Research) and SIRENE (*Système national d'identification et du répertoire des entreprises et de leurs établissements,* referential for French corporations maintained by INSEE) for affiliations. Except for ORCID, (HAL to some extent), and grid, the other databases and referential mentioned here are nation-specific. In particular, for authors identification, we really leverage on the IDRef referential.

19  In this paper, we do not cover the identification methods for authors and structures and leave it for future work.

### Scientific field classification

20  We want to estimate the Open Access prevalence by discipline. The task of estimating what is the scientific discipline of a given publication could leverage on the content of the publication (title, keywords, summary and full text), the journal in which it is published, and also on the authors themselves.

21  In this work, we use only "content data", trying to infer information from the title of the publication. We focus on the title as it is the only content data available at a large-scale.

22  Clustering a set of documents by topic is a difficult Natural Language Processing (NLP) task (unsupervised learning). Instead, we chose to use a supervised learning method, but we needed a training dataset.

23  We used the "Pascal-Francis" open database from INIST, CNRS, with 17M+ publications. This training dataset labels each of the 17 million entries with tags from a hierarchical classification of a few thousand tags.

24  We then train a model to predict, from a title, the Pascal-Francis classification tags. To do so, we used the approach developed by Joulin, Grave et al (2017) called FastText classification. It leverages on word embeddings, including bi-grams.

25  The model was evaluated with a train-test split of 80% - 20% and then trained on the whole dataset before being applied to predict scientific field on new data.

26  Once this model built, we then cluster the Pascal-Francis tags into 10 macro-disciplines, namely: Biology, Engineering, Social sciences, Mathematics, Computer and information sciences, Chemistry, Medical research, Physical sciences / Astronomy, Earth / Ecology / Energy / Applied biology, Humanities.

27  We add an extra step in the scientific field classification when the associated probability given by the model is too low. That means the FastText classifier is really unsure of the predicted tags. That can be the case for example when the title is too short. There, scientific discipline is consolidated at the journal level. So publications that have no scientific discipline get the scientific discipline that is the most common in the same journal.

## Third step in the method: Open Access detection

28  Eventually, we detect if a publication is OA using data from Unpaywall and HAL. We fetch not only the OA status but also information about licensing and hosting.

29  The main point in this third step is actually the dynamics. That is to say that the OA status of a publication evolves over time. In particular because of the embargo period for open repositories and moving barriers for journals. In this work, we have used cold data from the June 2018 snapshot of the Unpaywall database. However, we plan in future work to keep updating the data with warm data coming from Unpaywall data feed. A shortfall here of the Unpaywall data is that it does not contain the date of the first time a publication was detected as OA, but instead the last time. Memorizing what is the first time the publication was OA would enable to analyze how long it takes, after the publication date, for a publication to become OA. This will be part of our further work.

Monitoring Open Access at a national level: French case study    530  Also, different types of OA exists. Piwowar, Priem et al (2018) differentiate several types: gold, green, hybrid and bronze. In this study, for a matter of simplicity, we only differentiate publisher hosted OA and repository hosted OA.

## RESULTS

31  All the following results were based on the Unpaywall snapshot from June 2018. Unpaywall also provides a feed dataflow with weekly updates that we have not yet integrated into this work.

### How many publications with a French affiliation are identified?

32  For the year 2017, we identify about 130,000 publications with a French affiliation. Over the 5-year period, results are given in Table 1. It is difficult to benchmark these results as there is no clear number of the total publishing output in France. Still, for instance, for the year 2017, we identify 133k publications, whereas around 110k publications are identified for France in the year 2017 in the Web of Science.

Table 1: number of publications detected with a French affiliation year by year

| Publication year | Number of publications detected with a French affiliation |
| --- | --- |
| 2013 | 114, 424 |
| 2014 | 117, 900 |
| 2015 | 116, 722 |
| 2016 | 127, 470 |
| 2017 | 132, 970 |

### What is the precision of the identification method?

33  We estimate the precision of this detection with a manual check on a random sample of 100 publications for the year 2017.
With this manual check, we estimate the precision of the method at 96% (4% of false positive error). Instead, the recall (proportion of publication with a French affiliation that we did not detect) is much more difficult to measure, as the French production (around 100k publications a year) represents a small fraction of the global scientific production.

### What are the main types of publications for the identified DOIs?

34  The majority of publications that we analyze are journal-articles (86.7%). Books and monographs only represent 0.5% of the total for the year 2017. That is probably an underestimation of the reality. It is actually a consequence of our first choice to reduce the perimeter to publications with a DOI.

ELPUB 2019

Table 2: number of publications detected split by type, for the year 2017

| Publication type | Absolute number (2017) | Percentage (2017) |
|---|---|---|
| journal-article | 115,301 | 86,7% |
| book-chapter | 8,681 | 6,5% |
| proceedings-article | 7,004 | 5,3% |
| book / monograph | 646 | 0,5% |
| other | 1,338 | 1% |

## What are the main publishers for the identified DOIs?

35  Elsevier BV represents about 30% of the total. Two publishers specialized in social sciences and humanities, mainly in French language (CAIRN and OpenEdition) represent put together 5.5% of the total.

Figure 1: Part of the top-10 publishers (in percentage of the all the publications detected), for the year 2017

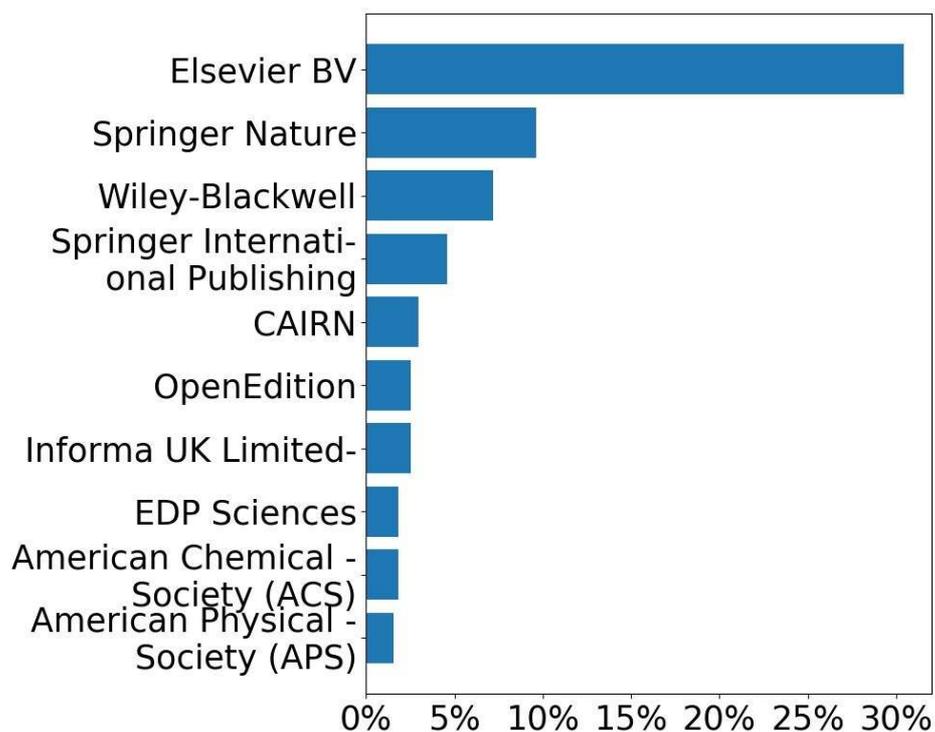

## What is the split of the identified DOIs per discipline?

36  The scientific discipline of a publication is not part of its available metadata. So we built a machine-learning classifier that has first to be evaluated.

37  We evaluate the performance of the classification model with two metrics, precision and recall. The precision is the number of correct labels among the labels predicted by the model. Conversely, the recall is the number of labels that successfully were predicted, among all the real labels. The real labels in our case are the Pascal-Francis labels in the training set.

38  As an entry can have multiple labels in the training dataset, we evaluated the model in two contexts, with only the first label predicted (Precision@1 and Recall@1), and with the first five labels predicted (Precision@5 and Recall@5). The FastText tutorial webpage gives more detail on these metrics.

Table 3: performance on the testing set of the calibration model

| Performance Metric | Value |
|---|---|
| Precision@1 | 86% |
| Recall@1 | 13% |
| Precision@5 | 60% |
| Recall@5 | 47% |

39  We applied this classifier and clustered its tags into 10 macro disciplines and get the discipline split-up for 2017 identified DOIs in Figure 2.
As expected Medical research has the most important share in the number of publications. We also note that the Social sciences and Humanities put together gather 10% of the total number of identified DOIs for 2017. That could be an underestimate as publications without a DOI are excluded from the method presented (in particular the share of books is low, as shown in Table 2).

Figure 2: Scientific field repartition of the identified DOIs for the year 2017

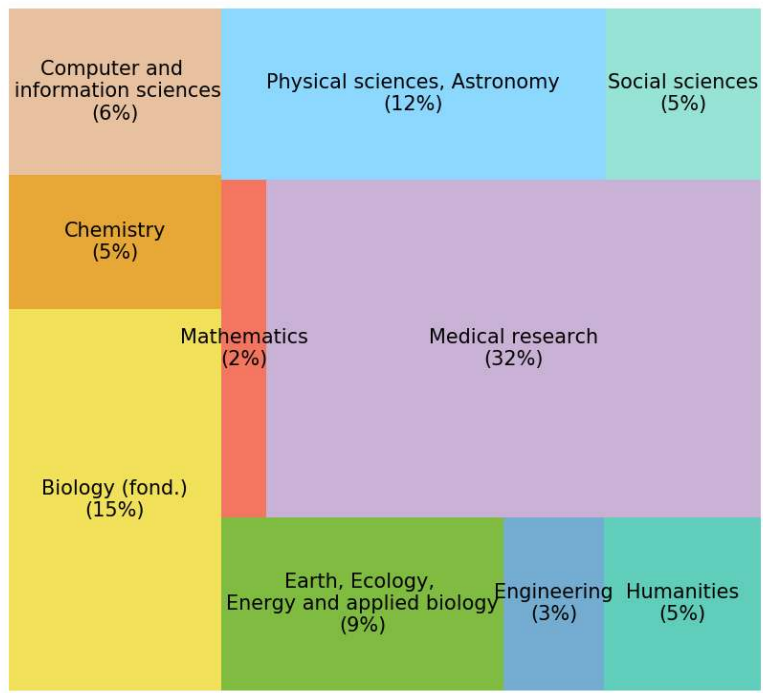

## What is the precision of Open Access detection?

40  For Open Access, the method described above entirely relies on Unpaywall and HAL. So first, we need to check how reliable the sources are.
Indeed, Akbaritabar and Stahlschmidt (2019) found 15% of false negative, i.e publications indicated as closed whereas they actually are open access.
Also, Mikki, Susanne, et al. (2018) show that, for the Norwegian case, the national OA rate measured with oaDOI (previous version of Unpaywall) is 31% when they state it is actually at 70%.

41  We manually check a random sample of publications to assess their open-access status. Given a publication marked as closed in Unpaywall snapshot we used, three cases can happen:
- It is still closed access as of the time of the manual check.
- It is open access, and it is also seen as open access in the Unpaywall feed data (data has evolved since the snapshot we used)
- It is open access, but still seen as closed in the Unpaywall feed data.

42  With two random samples of publications (of size 100 each), one for the year 2017, and one for the year 2013, we get the following results:

Table 4: evaluation of the June 18 Unpaywall snapshot recall for OA detection (DOI marked closed access while it is open access), as of March 2019.

| Publication year | % of false negative, now marked OA in Unpaywall feed update | % of false negative, still marked closed in Unpaywall feed update | Total false negative |
|---|---|---|---|
| 2013 | 2% | 1% | 3% |
| 2017 | 5% | 6% | 11% |

43   These results show that the OA status is hard to get right, and moving in time. Even for many-year-old publications (e.g from 2013), we estimate the false negative rate at 3%. This figure can be decomposed in two parts: 2% that were marked as closed in the Unpaywall data we used (from June 18) but now marked as open in the updates; and 1% still marked as closed.

44   Two hypotheses could explain the first part: either the OA status has actually changed, more than 6 years after publication; either Unpaywall engine got improved in the meantime and now catches DOI that it did not catch a few months before. We do not have enough evidence to decide what is the right explanation.

45   This phenomenon gets amplified for more recent publications. Their OA status actually moves for sure, because of the embargo and moving barriers.
We remark that most of the false negative cases are publisher-hosted open access.

## What is the overall level of Open Access on the period 2013-2017?

46   We observe in Figure 3 a gentle increase from 39.6% in 2013 to 42.7% in 2016 of estimated OA rate for detected French publications. In 2017, we see a small decrease to 40.5%. However, we analyze this decrease as a consequence of the difficulty to catch the OA status dynamics underlined above, and not as a real decreasing trend.
Moreover, this decrease is backed only by publisher hosted OA, which is in line with our previous observation that most of the false negative cases are publisher hosted. These OA rate figures should then be seen as a snapshot of the situation as of June 2018 (date of the Unpaywall snapshot).

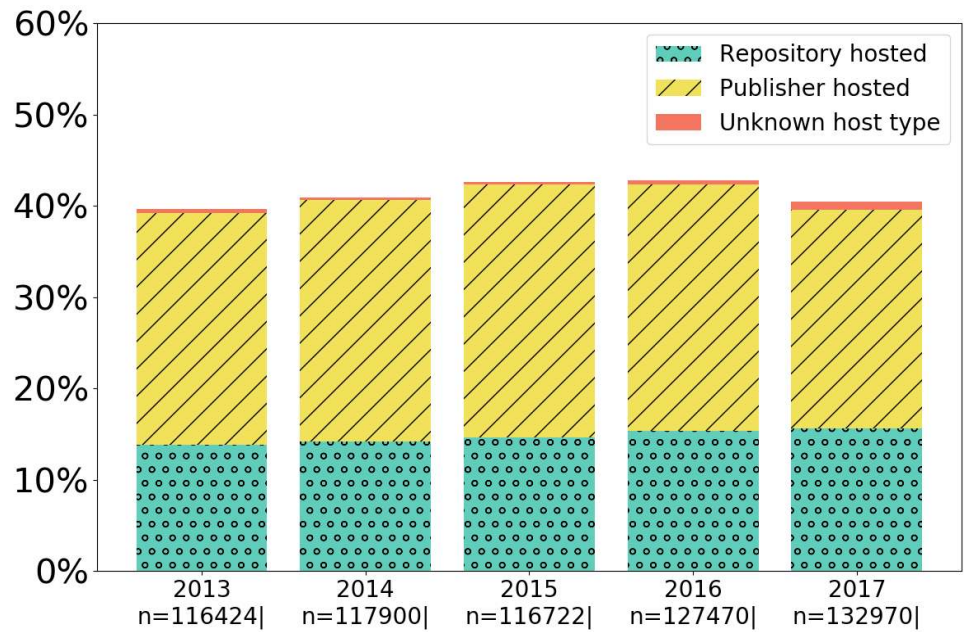

Figure 3: Evolution of the OA rate in detected French publications, from 2013 to 2017 (provisional figures)

47  They will actually evolve with time, and be affected by multiple factors:
    - OA status does change over time
    - Unpaywall engine should hopefully keep improving over time, reducing false-negative rate
    - DOIs are created for past publications (DOI backfill). So the underlying numbers of identified DOIs could themselves evolve.

48  As a consequence, the database we build with our method should be considered as a living database, that keeps updating from new pieces of information (new DOIs, new OA status), even for objects whose publication date is several years back in the past.
    In order to analyze these dynamics, it will be key to also memorize the first date of observed Open Access. That will allow measuring the amount of time it gets, after publication, to actually observe it has become OA.

49  Note that this amount of time will actually merge the actual time to become OA (embargo, moving barrier, etc.) and also the observation lag (i.e after a publication really becomes OA, it may take some time to observe this transition).

## What is the Open Access level per type of publication (year 2017)?

50  For the year 2017, we observe in Figure 4 an OA rate of 42% for journal-article, 39% for proceedings article, 17% for book-chapters and books.
    In terms of hosting, the majority of OA journal-article are publisher hosted while it is the other way around for proceedings-article.

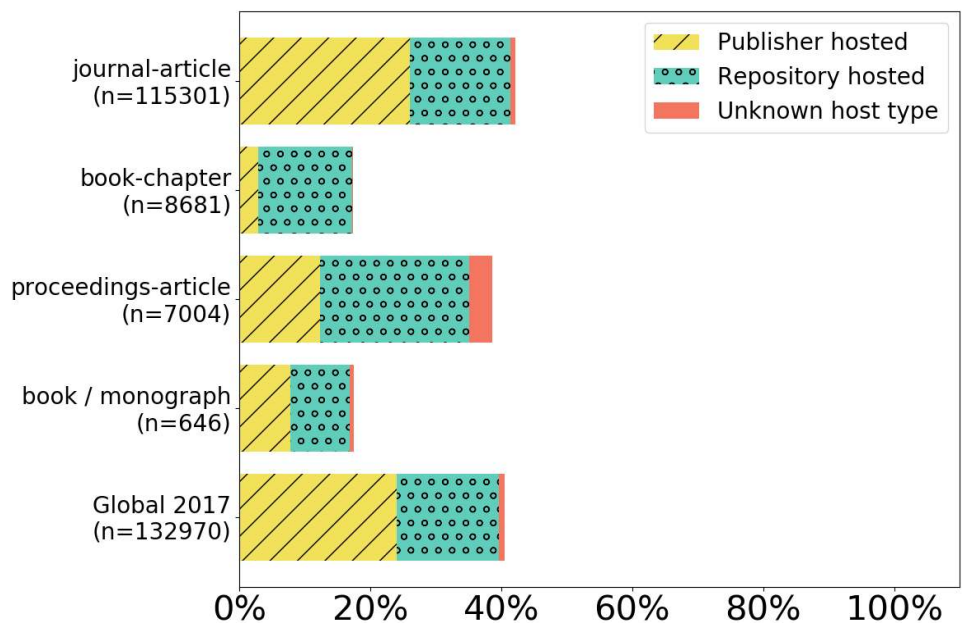

Figure 4: OA rate per publication type in 2017, for detected French affiliations

## What is the Open Access level per discipline (year 2017)?

51  Figure 5 shows that OA penetration is quite different from a discipline to another. The OA rate ranges from more than 60% for Mathematics to lower than 35% for several disciplines, such as Engineering, Humanities, Chemistry, Medical research and Social sciences.

In particular, we notice that the first discipline in terms of production, Medical research (about a third of the total number of identified publications) has one of the lowest OA penetration rate.

We also observe that the share of publisher hosted OA versus repository hosted OA is very different from a discipline to another. Indeed, the high OA rate in Mathematics is driven by repository hosted OA while it is the other way around for Biology (fundamental).

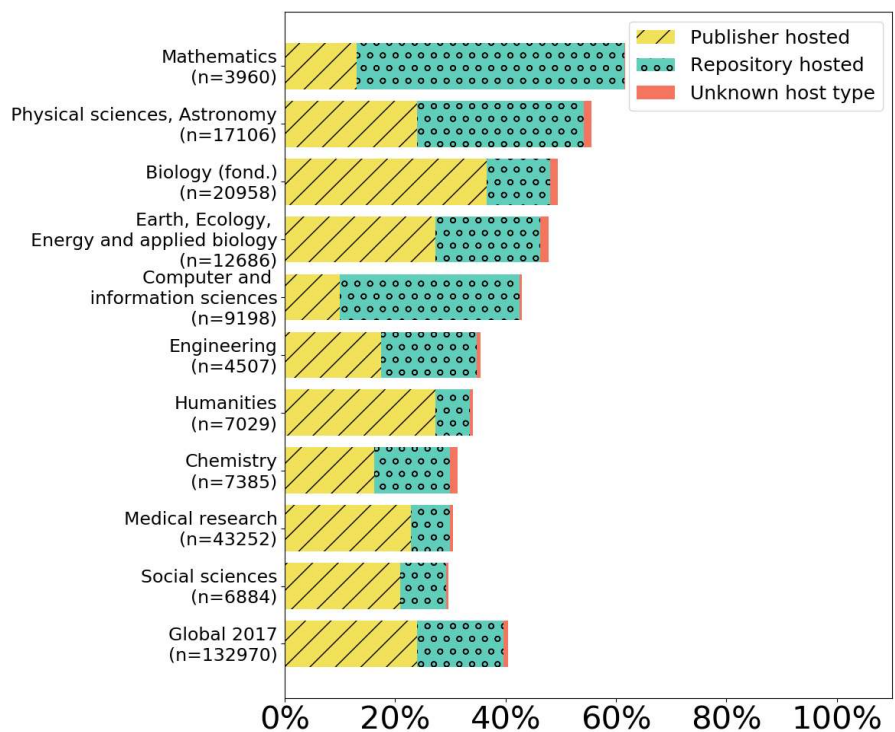

Figure 5: OA rate per publication type in 2017, for the detected French affiliations

## What is the Open Access level per publisher (top-10 publishers for 2017)?

52  Again, when it comes to OA penetration per publisher, we observe large variations from a publisher to another (Figure 6).

53  These variations could be explained with several factors. First, of course, it will depend on the publisher's policy towards open access. It is also coupled with large variations across disciplines. That is what we observe when we look at some discipline-specific publisher. For example, the gap observed between the ACS (30% of OA) and the APS (84% of OA) for the identified DOIs in 2017 is in line with the gap observed at the discipline level between Chemistry and Physics.

Figure 6: OA rate per publishers in 2017, for the top-10 publishers of the detected French affiliations

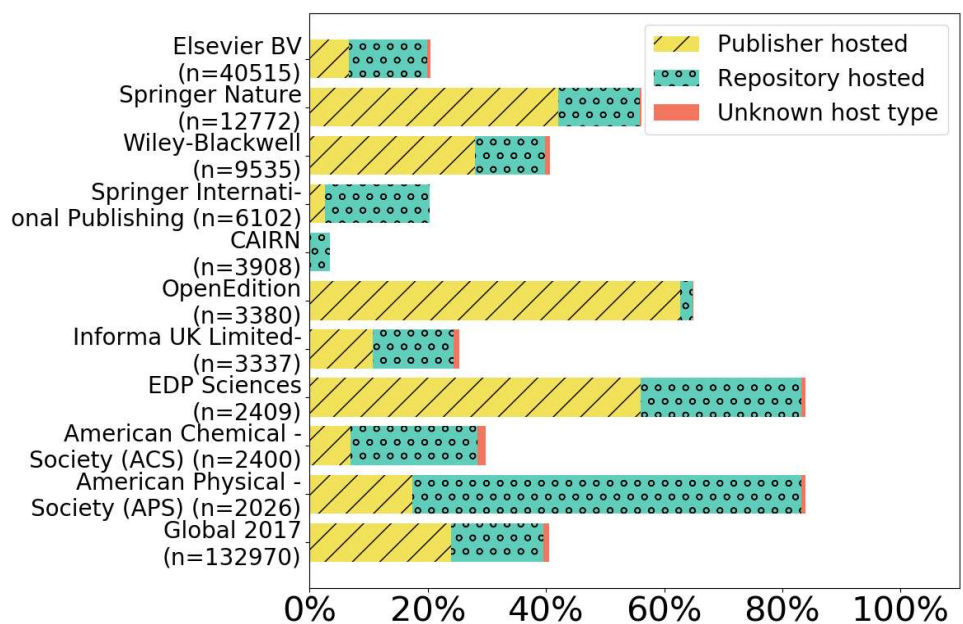

## Discussion and Conclusion

54  As public policies towards Open Access are developing, measuring Open Access has become a need for public policymakers.
We built a methodology that does not rely on any commercial database and that is transparent. Our implementation is open sourced on this Github repository https://github.com/dataesr/publications and the resulting database will be set as open data. This work addresses the following research question: what percent of the French scientific production is OA and how does it vary with discipline, publisher and with time?

55  We first identified a set of publications as a working perimeter, that we will consider for this study as the French publications on the period.
We found that the global OA rate on these publications increased from 39.6% in 2013 to 42.7% in 2016. We observe a slight decrease for the last observed year (2017), but we analyze this as a consequence of the fast-moving nature of OA status. This means that this measure is probably an underestimate of the current OA rate for 2017 in France.
As a consequence, a deep study of the dynamics of the OA status remains to be done.

56  We also found the OA rate variations are huge from a discipline to another. The results presented in Figure 5 remain raw figures. It would be very interesting to explain those with the actual uses in these disciplines. Moreover, lots of factors could impact these differences. For example, the funding level is not the same between discipline. Nor are the journal prices from a discipline to another. That leaves room for future research to understand these differences.

57  Another key point we observe is that overall about 15% of the publications are green open access (repository hosted). That could seem low. This is actually underestimated in this analysis as the publications that are both available on the journal website and also on a repository is only counted as publisher hosted OA. But there is still much room for improvement for green OA. Indeed, with the evolution of the French law (Digital Republic Act) in October 2016, authors whose research has been publicly funded at more than 50% have the right (not the obligation) to post their work on a repository, after an embargo period of maximum 12 months (depending on the disciplines). So we expect the share of repository hosted OA to grow faster in the next years.

58  This work also presents several limits.
First, we included in the analysis only the publications with a DOI. That excludes from the analysis a share of the production, in particular, a part of books.
Moreover, to compensate for the metadata poverty, we add to enrich them: French affiliation detection, author identification, discipline classification. Each of these enrichments is automatic, based on rules (regex like or heuristics) and machine learning. So there are errors, that we tried to estimate.
Another limit of the analysis is the difficulty to get the OA status right. In particular because of its dynamics, and also because we rely on the Unpaywall data that is itself containing errors, in particular, false negative.

59  The design and the implementation of this methodology is transparent and independent from commercial databases.
Some bits of the method are specific to the French case. It concerns the alignment with French referential, like IDref for authors. ORCID is a good international identifier, but in our case, we rely on the IDRef producer (ABES) that already automatically match ORCID and IDRef once a year. But parts of it are totally generic: identifying the publication for a given perimeter (country, region...) is based only on rules that can be changed; the discipline classification is generic (but trained on the Pascal-Francis database); the OA status comes from Unpaywall which is a large-scale database.

60  So we hope other countries will help develop this kind of methodologies and tools.

## FURTHER WORK

61  This work is only a first step to build a large-scale, open, database for Open Access monitoring.

62  Upcoming work is about authors and affiliations identification in several external referential, especially IdRef, ORCID, GRID, RNSR, SIRENE, HAL to improve data interoperability. That means handling the disambiguation for authors and affiliations, and use existing, external identifiers for these objects.

63  Improving the detection of the publications with a French affiliation is also part of the next steps. Especially how we could catch publications without DOI.

64  Besides, a large piece of work concerns the measure and the understanding of the OA status dynamics. At what pace does the OA status move?
Next steps will also include the addition of new data sources, namely OpenAPC for Article Processing Charges (APC) monitoring and OpenCitations to add references and citations.

## ABSTRACT

After the launch of multiple plans for Open Science, there is now a need for an accurate method or tool to monitor the Open Science trends and in particular Open Access (OA) trends. We address this requirement with a methodology that we developed and tested for France, but that could be extended to other countries. Only open data and information available on the Web are used, leveraging as much as we can large-scale systems such as Unpaywall, HAL (the main open repository in France, part of the CNRS), ORCID and IDRef (referential for French Higher Education and Research). We used rule-based and machine learning techniques to enrich the metadata of the publications. We estimate that the overall OA rate for French affiliated publications ranges from 39% to 42% between 2013 and 2017. The trend is slightly up, except for the last year, but we gather evidence that shows this is a consequence of the moving nature of the OA status. Therefore these figures should be seen as a snapshot rather than definitive. For the last observed year (2017), we show that the OA rate varies according to the publication type, the publisher and the discipline (more than 60% in Mathematics while it is about 30% in Medical research which represents the largest share in the number of publications). We describe the main challenges of our method (detection of the publications with a French affiliation, metadata enrichment with machine learning, open access status) and evaluate the errors of each step. Most of the method is not country-specific and could be applied for another perimeter. Our implementation is open sourced on the repository https://github.com/dataesr/publications.

## INDEX

## AUTHOR

ERIC JEANGIRARD

Ministry of Higher Education, Research and Innovation, Paris, France
eric.jeangirard [at] recherche.gouv.fr